\documentclass[preprint]{aastex}

\newcommand{\HI}{H{\,\small I}}

\def\NII{[\ion{N}{2}] $\lambda\lambda6548,84$}

\def\OIII{[O{\,\small III}]}
\def\NII{[N{\,\small II}]}

\newcommand{\kms}{km s$^{-1}$}

\input psfig

\slugcomment{Submitted to AJ, \today}

\shorttitle{VLBI Observations of the Seyfert galaxy IC~5063}
\shortauthors{Oosterloo et al.}
\begin{document}

\title{A strong jet/cloud interaction
in the Seyfert galaxy IC~5063: VLBI observations}

\author{T.A. Oosterloo, R. Morganti\altaffilmark{1},}
\affil{Netherlands Foundation for
Research in Astronomy, Postbus 2, 7990 AA, Dwingeloo, NL}
\email{oosterloo@nfra.nl}
\email{morganti@nfra.nl}

\author{A. Tzioumis, J. Reynolds, E. King,}
\affil{ATNF-CSIRO, P.O. Box 76,
Epping NSW 1710, Australia}

\author{P. McCulloch,} 
\affil{Dept. of Physics, University of Tasmania,
GPO Box 252-21, Hobart Tas 7001, Australia}

\and

\author{Z. Tsvetanov}
\affil{ Johns Hopkins University, 3400 North Charles Street, Baltimore, MD 21218,
USA}

\altaffiltext{1}{Istituto di Radioastronomia, CNR, via Gobetti 101, 40129 Bologna, Italy}

\begin{abstract} 

We present 21-cm \HI\ line and 13-cm continuum observations,
obtained with the Australian Long Baseline Array, of the Seyfert 2 galaxy
IC~5063.  This object appears to be one of the best examples of Seyfert
galaxies where shocks produced by the radio plasma jet influence both the
radio as well as the near-infrared emission.  The picture resulting from the
new observations of IC~5063 confirms and completes the one derived from
previous Australia Telescope Compact Array (ATCA) lower resolution
observations.  A strong interaction between the radio plasma ejected from the
nucleus and a molecular cloud of the ISM is occurring at the position of the
western hot spot, about 0.6~kpc from the active nucleus.  Because of this
interaction, the gas is swept up forming, around the radio lobe, a cocoon-like
structure where the gas is moving at high speed.  Due to this, part of the
molecular gas is dissociated and becomes neutral or even ionised if the UV
continuum produced by the shocks is hard and powerful enough.

In the 21-cm \HI\ line new data, we detect only part of the strong
blue-shifted \HI\ absorption that was previously observed with the ATCA at
lower resolution.  In particular, the main component detected in the VLBI
absorption profile corresponds to the most blue-shifted component in the ATCA
data, with a central velocity of 2786 \kms\ and therefore blue-shifted
$\sim$614 \kms\ with respect to the systemic velocity.  Its peak optical depth
is 5.4\%.  The corresponding column density of the detected absorption, for a
spin temperature of 100 K, is $N_{\rm HI} \sim 2 \times 10^{21} $atoms
cm$^{-2}$.  Most of the remaining blue-shifted components detected in the ATCA
\HI\ absorption profile are now undetected, presumably because this absorption
occurs against continuum emission that is resolved out in these
high-resolution observations. 

 The \HI\ absorption properties observed in IC 5063 appear different from those
observed in other Seyfert galaxies, where the \HI\ absorption detected is
attributed to undisturbed foreground gas associated  with the
large-scale  galaxy disk. In the case of IC~5063, only a small fraction
of the absorption can perhaps be due to this.  The reason for this can be that
the western jet in IC~5063 passes through a particularly rich ISM.
Alternatively, because of the relatively strong radio flux produced by this
strong interaction, and the high spectral dynamic range of our observations,
broad absorption lines of low optical depth as detected in IC 5063 may have
remained undetected in other Seyferts that are typically much weaker radio
emitters or for which existing data is of poorer quality.

\end{abstract}


\section{Introduction}

The study of the effects of interactions between the radio plasma ejected from
an active nucleus and the interstellar medium (ISM) of the hosting galaxy is
presently attracting a lot of interest.  In particular, Seyfert and
high-redshift radio galaxies appear to be the kind of objects where the
effects of such interactions can be very important.  They can range from
shaping the morphology of the gas in the ISM (with the radio plasma sweeping
up material as it advances in the ISM), to the ionisation of the gas itself. 
While there is little doubt on the presence of such interactions in objects
like Seyferts or high-$z$ radio galaxies, the actual importance of these
effects in determining the overall characteristics of these sources is still
a matter of debate. 

In some Seyfert galaxies the morphological association between the radio
plasma and the optical line-emitting clouds, as well as the presence of
disturbed kinematics in these clouds, is striking.  In particular, the
narrow-line regions (NLR) in Seyfert galaxies (i.e.\ regions of highly
ionised, kinematically complicated gas emission that occupy the central area
-- up to $\sim 1$ kpc from the nucleus) often appears to form a `cocoon'
around the radio continuum emission (see e.g.\ Wilson 1997 for a review;
Capetti et al.\ 1996; Falcke, Wilson \& Simpson (1998) and references
therein). Moreover, outflow phenomena are observed in the warm gas of several
Seyfert galaxies (see Aoki et al.\ 1996 for a summary).  Thus, the NLRs
represent some of the best examples of regions where interaction between the
local ISM and the radio plasma takes place and can be studied in detail.

The situation appears to be different for the atomic hydrogen.  Observations
of the \HI\ 21-cm line, in absorption, can trace the distribution of this gas
in front of the brightest radio components, that are usually observed in the
central region of Seyferts (of kpc or sub-kpc size, i.e.\ {\sl co-spatial with
the NLRs}).  Thus, the study of the distribution and kinematics of the {\sl
cold} component of the circumnuclear ISM can nicely complement the optical
data.  Although \HI\ absorption has been detected in a number of Seyfert
galaxies (e.g.\ NGC~4151 Pedlar et al.\ 1992; NGC~5929, Cole et al.\ 1998;
Mkn~6, Gallimore et al.\ 1998 ; see also Brinks \& Mundell 1996 and Gallimore
et al.\ 1999 and references therein), most of the investigated objects show
single localised \HI\ absorption components that can be explained as rotating,
inclined disks or rings aligned with the outer galaxy disk (Gallimore et al.\
1999) and only very seldom with gas in a parsec-scale circumnuclear torus
(NGC~4151, Mundell et al.\ 1995).  These components are therefore originated
by gas that is not in interaction with the radio plasma. 

However, more complex {\sc H\,i} absorption profiles that cannot be explained
by the above mechanism have been observed in at least one Seyfert galaxy,
IC~5063.  Australia Telescope Compact Array (ATCA) observations of this galaxy
(Morganti, Oosterloo \& Tsvetanov 1998, hereafter M98) have revealed a very interesting
absorption system with velocities up to $\sim 700$ \kms\ blue-shifted with
respect to the systemic velocity.  In this object, unlike in other Seyfert
galaxies, at least some of the observed \HI\ absorption is originating from
regions of interaction between the radio plasma and the ISM, producing an
outflow of the neutral gas.  
This object, therefore, poses a number of interesting questions as:
where is the interaction occurring, what are the physical conditions, why such
interaction is not  seen more often in neutral gas in other Seyfert galaxies?

Previous \HI\ observations were limited by low spatial resolution. In this
paper  we
present the results from new VLBI observations aimed at investigating in more
detail its nuclear radio structure and locating where the complex \HI\ absorption
observed with ATCA is really occurring.

Throughout the paper we adopt a Hubble constant of $H_\circ = 50$ \kms, so that
1~arcsec corresponds to 0.32 kpc at the redshift of IC~5063.

\section{Summary of the properties of  IC~5063}

IC~5063 is a nearby ($z = 0.0110$) early-type galaxy that hosts a Seyfert 2
nucleus that emits particularly strong at radio wavelengths ($P_{\rm 1.4\,GHz}
= 6.3\times 10^{23}$ W Hz$^{-1}$).  This object has been recently studied,
both in radio continuum at 8 GHz and in the 21-cm line of {\sc H\,i}, using
the ATCA (M98).  In the
continuum, on the arcsecond scale, we find a linear triple structure (see
Fig.\ 1) of about 4 arcsec size ($\sim 1.3$ kpc), that shows a close spatial
correlation with the optical ionised gas, very similar in nature to what is
observed in several other Seyfert galaxies (see e.g.\ Wilson 1997) and
indicating that the radio plasma is important in shaping the NLR.

In the \HI\ 21-cm line, apart from detecting the emission from the large-scale
disk of IC 5063, very broad ($\sim 700$ km\,s$^{-1}$), mainly blue-shifted
absorption was detected against the central continuum source. These line
observations could only be obtained with $\sim 7$ arcsec resolution, the
highest resolution achievable with the ATCA at this wavelength. This
resolution is too low to resolve the linear continuum structure detected in
the 8-GHz continuum image.  However, and what makes this absorption
particularly interesting is that we were able to conclude (by a careful
analysis of the data, see M98 for the detailed discussion)
that at least the most blue-shifted absorption is likely to originate against
the western (and brighter) radio knot and not against the central radio
feature seen at 8 GHz.  The large, blue-shifted velocities observed in the
absorption profile make it very unlikely that these motions have a
gravitational origin (the most blue-shifted \HI\ emission associated with the
large-scale \HI\ disk occurs at roughly $300$ \kms with respect to the
systemic velocity), and are more likely to be connected to a fast outflow of
the ISM caused by an interaction with the radio plasma.

The identification of the central radio feature as the core, and hence that
the absorption is occurring against the western lobe, is an important element
in interpreting the nature of the absorption detected in IC~5063.  In the
literature, the core of this galaxy has sometimes been identified with the
bright western knot (Bransford et al.\ 1998), however in our opinion there is
compelling evidence that the identification of M98 is
correct.

The superposition of the 8-GHz radio image with an optical {\sl WFPC2} image  available from the
{\sl HST} public archive and with a ground based narrow-band \OIII\ image
suggests that the nucleus coincides with the central radio knot (see Figs 3
and 4 from M98).  Although, as usual, there is some freedom
in aligning the {\sl WFPC2} image with the 8-GHz radio image, aligning the
western radio knot with the nucleus would require too large a shift.  Given
that the {\sl WFPC2} image was taken through the F606W filter, it contains the
bright emission lines of \OIII$\lambda 5007$, H$\alpha$ and \NII
$\lambda\lambda 6548, 6584$, and it gives a good idea of the morphology of the
ionised gas.  By aligning the nucleus with the central radio knot, a good
overall correspondence between the radio morphology and the bright region of
optical emission lines is obtained, both in the {\sl WFPC2} image and the
ground-based \OIII\ image, similar in nature to what is observed in many other Seyfert
galaxies.  Choosing this alignment, the western radio knot falls right on top
of a very bright, unresolved, spike in the {\sl WFPC2} image, i.e.\ the western
radio knot would also have a counterpart in the {\sl WFPC2} image.  The
filamentary morphology of the ionised gas of the region just around this spike
is suggestive of an interaction between the radio plasma and the ISM and the
identification of the western radio lobe with this feature seems natural. 
Using optical spectroscopy, Wagner \& Appenzeller (1989) found off-centre
blue-shifted broad emission lines with similar widths as the detected \HI\
absorption at a position 1-2 arcsec west of the nucleus, i.e.\ coincident with
the spike seen in the {\sl WFPC2} image.  This also suggests that at this
position a violent interaction is occurring. 

The identification of the core with the central radio knot has been recently
confirmed by Kulkarni et al.\ (1998) from NICMOS images.  Three well resolved
knots were detected in the emission lines of [Fe\,{\small II}], Pa$\alpha$ and
H$_2$.  This emission-line structure shows a direct correspondence with the
radio continuum structure.  In broad band near IR images they detected a very red
point source coincident with the central source seen in the emission lines,
consistent with previous suggestions of a dust-obscured active nucleus.  The
strong [Fe\,{\small II}] and H$_2$ emission are usually interpreted as
evidence for fast shocks and the direct correspondence between these regions
and the radio emission suggest that shocks associated with the radio jet play
a role in the excitation of the emission-line knot. 

By the same authors, an asymmetry in the H$_2$ distribution was found, with the
eastern lobe showing a much weaker emission than the western lobe.  This
asymmetry can be explained, e.g, if an excess of molecular gas is present on
the western side (for example, if the radio jet has struck a molecular cloud).

In the optical, IC5063 shows a very high-excitation emission line spectrum
(including [Fe\,{\small VII}]$\lambda\lambda$5721, 6087; Colina, Spark \&
Macchetto 1991). The high-excitation lines are detected within 1 - 1.5 arcsec
on both sides of the nucleus, about the distance between the radio core and
both the lobes.  These lines indicate the presence of a powerful and hard
ionising continuum in the general area of the nucleus and the radio knots in IC~5063.  We have estimated
(M98) the energy flux in the radio plasma to be an order of
magnitude smaller than the energy flux emitted in emission lines.  The shocks
associated with the jet-ISM interaction are, therefore, unlikely to account
for the overall ionisation and the NLR must be, at least partly,
photoionised by the nucleus, unless the lobe plasma contains a significant
thermal component (Bicknell et al.\ 1998).

\section{VLBI observations}

IC 5063 was observed with the Australian Long Baseline Array (LBA) initially in
continuum at 13 cm (2.3 GHz), followed by spectral-line observations at the
frequency corresponding to the redshifted \HI. 

The 13-cm observations in June 1996 comprised five stations; Parkes (64 m),
Mopra (22 m), the Australia Telescope Compact Array (5$\times$22-m dishes as
tied-array), the Mount Pleasant 26-m antenna of the University of Tasmania and
the Tidbinbilla 70-m antenna of the Canberra Deep Space Communications Complex
(CDSCC) near Canberra.  The observations used the S2 recording system to
record a single 16 MHz band in right-circular polarisation and were correlated
at the LBA S2 VLBI correlator of the Australia Telescope National Facility at
Marsfield, Sydney. 

The 13-cm data were edited and calibrated using the AIPS processing system.  After
this, the data were exported to DIFMAP (Sheperd 1997) for model fitting and
imaging.  The final image is presented in Fig.~2 and was made with uniform
weighting. 

Although the observations were not phase-referenced, absolute position   
calibration for the 13-cm LBA image was extracted from the delay and rate
data, allowing the radio image to be fixed at the $\sim$0.1 arcsec level in   
each coordinate, adequate for registration with other images.

The 21-cm observations were made in September 1997 at the redshifted \HI\
frequency of 1407~MHz, recording 16~MHz bandwidths in each circular
polarisation. The same array was used, except for the Tidbinbilla 70-m
antenna, which has no 21~cm capability. Correlation was in spectral-line
mode with 256 spectral channels on each baseline and polarisation.

The editing and part of the calibration of the 21-cm line data was done in
AIPS and then the data were transfered to MIRIAD (Sault, Teuben \& Wright
1995) for the bandpass calibration.  The calibration of the bandpass was done
using additional observations of the strong calibrators PKS~1921--293 and
PKS~0537--441.

Problems were encountered at Mopra which limited the usefulness of those data.
It proved not possible to image the source from the final dataset and instead
a simpler analysis using the time-averaged baseline spectra was employed.

\section{The sub-kpc structure}

\subsection{The radio continuum morphology}

The final 13-cm image, shown in Fig.~2, has a beam of $\sim56\times 15\,$mas
in position angle (p.a.) $-40^\circ$.  The r.m.s.  noise is $\sim 0.7$ mJy
beam$^{-1}$.  The total flux is
210 mJy.  Because of the high accuracy of the astrometry of this VLBI image, we
know that the observed structure corresponds (as expected) to the brighter,
western, lobe observed in the 8-GHz ATCA image (see Fig.~2). 
It is therefore situated at about 0.6 kpc from the nucleus.

The image shows that the lobe appears to have a relatively bright peak ($77$
mJy beam$^{-1}$) and some extended emission to the north-east in p.a.~$\sim
40^\circ$ of total size of about 50 mas (or $\sim 16$ pc).  The p.a.\ is quite
different from the p.a.\ of the arcsecond sized structure seen in the ATCA
8-GHz data (p.a.~$\sim 295^\circ$), so there appears to be structure
perpendicular to the main radio structure.  These kind of distortions are
often seen in the radio structure of Seyfert galaxies (e.g.  Falcke et al. 
1998) and could perhaps result from the interaction of the radio plasma with
the environment. 

From our data, a brightness temperature of $\rm{T_{B}}\sim 10^{7}$K can be
inferred for the VLBI source.  This brightness temperature is several orders
of magnitude less than the typical values seen in milliarcsecond AGN cores or
inner (pc-scale) jets that typically have brightness temperatures between
$10^{9}$ and $10^{11}$K.  However, this temperature is quite commonly found
for radio knots detected in Seyfert galaxies (e.g.\ knot C in NGC~1068, Roy et
al.\ 1998).  Unfortunately, we do not have a spectral index of this region on
the VLBI scale.  The overall spectral index inferred from the ATCA 8.6 and
1.4-GHz images is steep, $\alpha\sim -1$, and indeed consistent with a radio
lobe or jet.  However, unless a detailed multi-frequency spectral index study
can be carried out, it is difficult to derive conclusions from this result
alone given the complexity often observed in the spectral index of the central
regions of Seyfert galaxies. 

In summary, we can conclude that the radio morphology, the spectrum and the
brightness temperature of the VLBI source are consistent with what expected in
a radio lobe.

\subsection{The \HI\ absorption}

As mentioned above, 
because from the 21-cm line observations useful data could only be obtained on
the Parkes-ATCA baseline, we will present only an time-integrated spectral
profile of the \HI\ on this baseline.  These data correspond to a spatial
scale of about 0.1 arcsec.

Fig.\ 3 shows the continuum-weighted \HI\ absorption profile.  
Heliocentric,
optical velocities are used.  For comparison, the spectrum obtained from the
previous ATCA observations (with much lower spatial resolution) is
superimposed (dashed line).  In Fig.\ 3 we have also indicated the range of
the velocities observed as measured for the \HI\ emission of the large-scale
disk of IC 5063, as well as the systemic velocity of the galaxy of $3400$ km
s$^{-1}$ as derived from the kinematics of the \HI\ emission.  The r.m.s.
limit to the optical depth is $\sim 0.3$\%.

Fig.\ 3 shows that a strong absorption signal is detected against the VLBI
source. Since from the 13-cm data it followed that the VLBI source corresponds
to the western radio lobe, these data now confirm what was believed to be the
case from the ATCA data, namely that the absorption is occurring against the
western radio lobe.
Fig.\ 4 shows the same data as in Fig.\ 3, except that both profiles have been
normalised to the same optical depth for the most blue-shifted component.

Figs 3 and 4 show quite clearly that the shape of the absorption profile
obtained at the high resolution of the VLBI data is quite different in
character than that obtained with the ATCA.  While in the ATCA data the
absorption is relatively uniform in velocity, in the VLBI spectrum the
most blue-shifted component is clearly the dominant one.  This shows
that the most blue-shifted absorption is occurring against a compact
radio source, while the absorption at lower velocities is against a more
diffuse source.  Component ($A$) has a central velocity of 2786 \kms,
over 600 \kms\ blue-shifted with respect to the systemic velocity (3400
\kms), with its bluest wing extending to about 2650 \kms, or --750 \kms\
relative to the systemic velocity.  Component $A$ corresponds to the
most blue-shifted component found in the ATCA profile, as is illustrated
in Fig.\ 4.  At slightly less blue-shifted velocities, but still outside
the range of velocities observed in emission, the VLBI data show a
second component ($B$).  The absorption with velocities within the range
of the \HI\ emission, as detected in the ATCA profile, is only partly
detected in the VLBI spectrum with component $C$.  No absorption is
detect in the velocity range 3000-3200 \kms.  Hence, the absorption seen
in the ATCA data at velocities above 3000 \kms\ has become much less
prominent compared to the more blue-shifted absorption.  Note that this
effect is probably even stronger than the data shows, since the low
resolution of the ATCA will have caused some filling of the absorption
with emission of the \HI\ disk and the `true' absorption is likely to be
stronger at these velocities.  The ATCA spectrum also showed a faint
red-shifted absorption component that is perhaps also detected in the
VLBI spectrum.

The column density $N_{\rm HI}$ of the obscuring neutral hydrogen is given by
$N_{\rm HI} = 1.823\times 10^{18} T_s \int \tau dv$ cm $^{-2}$ where $T_{\rm
s}$ is the spin temperature of the electron.  Assuming a spin temperature of
100~K we derive a column density of $\sim 1.7 \times 10^{21}$ atoms cm$^{-2}$
for the components $A$ and $B$ and a column density of $\sim 2.5 \times
10^{20}$ atoms cm$^{-2}$ for the component $C$.  The main source of
uncertainty for the derived column density comes from the assumption in the
value of the spin temperature.  The presence of a strong continuum source near
the \HI\ gas can make the radiative excitation of the \HI\ hyperfine state to
dominate over the, usually more important, collisional excitation (see e.g.\
Bahcall \& Ekers 1969).  Gallimore et al.\ (1999) argue that the \HI\ causing
the absorption against Seyferts jets is in general at too high densities
($\sim$$10^5$ cm$^{-3}$) for these effects to be relevant.  
However, the argument used by Gallimore et al.\ applies to \HI\ in pressure
equilibrium with the NLR, while the absorbing gas in Seyferts in general is
{\sl not} co-spatial with the NLR, but is at larger radii. Because of this the
density of the absorbing gas is lower, but the regions are also further
removed from the central engine and the spin temperature approaches the
kinetic temperature at much lower densities.
In our model for IC 5063, the \HI\ causing the most blue-shifted absorption
is the skin of a molecular cloud that is being stripped off  by the jet (see
also \S 5). Given that typical densities in molecular clouds are in the range
$10^4$ - $10^6$ cm$^{-3}$, this is an upper limit to the density of the
absorbing gas. But given the large velocities involved, the actual density of
the blue-shifted gas could be substantially lower.

The effects on the excitation of the fine-structure line by the local
radiation field were already discussed by M98 for the case
of IC 5063, where it was concluded that these effects are perhaps
important. The column density derived by Kulkarni et al.\  
from the NICMOS observation ($N_{\rm HI} \sim 5
\times 10^{21}$ atoms cm$^{-2}$) is slightly higher than our estimate based on 
a $T_{\rm spin}$ of 100 K, also suggesting that perhaps the spin temperature
is somewhat higher than 100 K. For $T_{\rm spin} = 100$ K the derived column
density is much lower than the value of $\sim10^{23}$atoms cm$^{-2}$ found
from X-ray data (Koyama et al.\ 1992).

\section{H{ \small \bf I}, H$_2$ and radio plasma: a possible scenario for the
interaction}

Summarising, the main result from our new observations is that with the
improved spatial resolution, the absorption at velocities outside the range
allowed by the rotational kinematics of the large-scale \HI\ disk has become
much stronger, while the absorption in the range of velocities of the \HI\
disk has become much less prominent. 

From the 13-cm radio continuum VLBI data we have been able to image only the
western part of the source observed by the ATCA, while the remaining structure
is resolved out.

Thus, all this confirms and completes the  picture we  derived from the ATCA data,
namely that {\sl a strong interaction between the radio plasma and the ISM is
occurring at the position of the western radio lobe}.  

Fig.\ 5 gives a schematic diagram of what we believe is happening in the western
lobe of IC~5063.  Following the results from NICMOS (Kulkarni et al.\  1998),
it is likely that the asymmetry observed in the brightness of the H$_2$
(western side brighter than the eastern side) may be explained by an excess of
molecular gas on the western side.  Thus, the radio plasma ejected from the
nucleus appears  to interact directly with such a molecular cloud.  Because
of this interaction, the jet is drilling a hole in the dense ISM, sweeping up
the gas and forming a cocoon-like structure
around the radio lobe where the gas is moving at high speed and an outflow of
gas is created.  The increased ultraviolet radiation due to the presence of
shocks generated from the interaction, can dissociate part of the molecular
gas.  This creates neutral hydrogen  or even ionised gas if the UV continuum
produced by the shocks is hard and powerful enough.  The region of ionised gas
would correspond to the part of the cocoon closer to where the interaction is
occurring, possibly corresponding to both the shocked gas and to the precursor.
The complex kinematics of the emission lines in this region observed from
the optical emission lines (Wagner \& Appenzeller 1989) is consistent with this.

As for the neutral gas, we will observe only the component in front of the
radio continuum and therefore, as effect of the outflow produced by the
interaction, we will observe only the blue-shifted component. The most
blue-shifted component will be seen against the hot spot where the interaction
is most intense. Somewhat away from this location, the \HI\ will  driven out by
the expanding cocoon, but since this is away from the hot spot, this will occur
at lower velocities. Moreover, the radio continuum emission from this region
is also more extended compared to the small-sized hot spot. Hence the VLBI
observation do not detect the absorption at lower velocities, but only the
highest velocities against the hot spot (as illustrated in Fig.5).

The origin of the H$_2$ emission can be related to UV or shocks (Draine,
Roberge \& Dalgano 1983, Sternberg \& Dalgano 1989).  Although we are not able
to distinguish between these two mechanisms, this scenario suggest that there
should be in IC~5063 a strong shock component.  The H$_2$ emission observed by NICMOS
would therefore come from the very dense region (again due to the compression
of the gas associated with the interaction) of the molecular cloud. 

As we noticed above, not all the components observed in the ATCA data are also
visible in the VLBI data.  A possible explanation for this is that the
components that are missing from the VLBI \HI\ absorption are against
continuum emission that is resolved out in our VLBI data, indicating that the
cocoon of shocked gas is quite extended and cover at least all the western
radio lobe.  Alternatively, part of the absorption undetected in the VLBI
spectrum can be due to the large-scale disk associated with the dust-lane and
also seen in \HI\ in emission, although the continuity of the ATCA absorption
profile does not suggest this.  

By looking at the velocity field derived for this disk from the \HI\ emission
observations (see Fig.\ 5 in M98) we can see as the western side
is the approaching side, therefore showing a blue-shifted velocity relative to
the systemic.  This means that the large-scale disk being in front of the hot
spot could be responsible for component $C$, but that it cannot explain the
weak red-shifted component unless non circular motions are present in the
foreground gas associated with the dust lane.  In M98 we
hypothesised that the red-shifted component could be associated with a nuclear
torus/disk.  This was motivated by the fact that the width of the red-shifted
component appeared to be similar to the width of the CO profile as observed by
Wiklind et al.\ (1995).  However, there seems to be no indication of detection
of the nuclear component from the visibility of the continuum associated with
the 21-cm data (and by extrapolating the arcsec data, the core flux is
probably too weak to be detected and, even more, to produce an absorption) so
this hypothesis has to be ruled out.  A final possibility is that, apart from
the bulk outflow, turbulent motions produced in the shocked region can give
rise to clouds with red-shifted velocity.

\section{Comparison with other Seyfert galaxies}

How does IC~5063 compare with other Seyferts galaxies?  The results on IC~5063
confirm the more general results obtained by Gallimore et al.\ (1999) on a
sample of Seyfert galaxies, that the \HI\ absorption is not occurring against
the core and that the absorbing gas in Seyferts does not trace (except for NGC
4151) the pc-scale gas. In the galaxies studied by Gallimore et al., the
absorption is occurring at a few hundred parsec from the core and is caused by
the inner regions of, or gas associated with, the large-scale \HI\ disks.
This is also happening in IC 5063.  The important difference is that in IC
5063 the jet is physically strongly interacting with this \HI\ disk, causing
the fast outflow observed for the absorbing material. This makes IC 5063
unique.  Only component $C$ would exactly fit the scenario proposed by
Gallimore et al. It is quite likely a gas cloud at large radius (given its
column density), unrelated to the interaction, projected in front of the radio
hot spot.

One obvious question is why such kind of absorption (i.e.\ broad blue-shifted
absorption) is so rare. Are the physical conditions in IC 5063 rare, or is
there an observational bias?

Some arguments suggest that IC 5063 is a special case.  IC 5063 is a very
strong radio emitter compared to other Seyfert galaxies.  Most of the strong
radio flux of IC 5063 is produced in the western radio knot, indicating that
the interaction is particularly strong.  Also the fact that the western radio
knot is much brighter than the eastern one indicates that the conditions near
the western lobe are special.  It has been noted that IC~5063 belongs to a
group of "radio-excess infrared galaxies" (Roy \& Norris 1997), objects that
could represent active galactic nuclei hosted in an unusual environment or
perhaps dust-enshrouded quasars or their progenitors. 
It appears that the jet-cloud interaction in IC 5063 is particularly strong. 
This would make IC 5063 a very suitable object for further detailed studies of
jet-cloud interactions in Seyfert galaxies. 

One factor is of course that in order to create the strong interaction and the
very broad absorption, the jet has to lie more or less in the plane of the
\HI\ disk.  Only then can the jet have a strong interaction with the ISM.  The
orientation of the AGN in Seyferts is not correlated with that of the
large-scale disk, so the effects seen in IC~5063 should then only occur in a
minority of cases. 

On the other hand, interactions between the radio plasma and the ISM are
common in Seyferts, given that in many Seyfert galaxies, very large velocity
widths of the optical emission lines are observed in the NLR (e.g.\ Aoki et
al.\ 1996 and references therein).  Perhaps the high sensitivity in $\tau$ of
our observations also plays a role.  IC~5063 is a strong radio source compared
to other Seyfert galaxies that are between 10 and 100 times weaker at radio
wavelengths.  Therefore only \HI\ absorption with much higher optical depth
can be observed against those objects. For example, the Seyfert 2 galaxy NGC
5929 shows a striking morphological similarity (both in the optical and radio)
with IC 5063.  However, the peak of the radio emission in NGC~5929 is only 24
mJy beam$^{-1}$, so in this object absorption of a few percent
would not be detectable with the noise level of the current observations (Cole
et al.\ 1998). For almost all the galaxies in the sample studied by Gallimore
et al.\ (1999) is the sensitivity not enough to have been able to detect
faint, broad absorption like in IC 5063. Moreover, in order to detect broad
profiles of the level as in IC 5063 even in strong sources, good spectral
dynamic range is required, which is not always easy to obtain (e.g.\ NGC 1068;
Gallimore et al.\ 1999). It is quite well possible that more cases like IC
5063 will be found if more sensitive observations are performed.

\section{Conclusions}

Using the Australian Long Baseline Array, we have detected a compact radio
source of about of 50 mas (or $\sim 16$ pc) in size (at 13 cm) in the Seyfert
galaxy IC 5063.  Because of the high positional accuracy of these
measurements, we can unambiguously identify this radio knot with the western
radio lobe.  The hot spot is extended in a direction almost perpendicular to
the radio jet. 

In 21-cm line observations, we detect absorption very much blue-shifted
($\sim$700 km s$^{-1}$) with respect to the systemic velocity.  Together with
the 13-cm observations, this confirms that the \HI\ absorption is not taking
place against the core, but that it is against the western radio knot.  At the
position of the western radio knot a very strong interaction must be occurring
between the radio jet and the ISM.  Various arguments suggest that this
interaction is particularly strong compared to other Seyfert galaxies.  This
makes IC 5063 a good candidate for studying the physics of jet-cloud
interactions in Seyfert galaxies. 

The HI absorption characteristics of IC 5063 are only partially
consistent with other absorption studies of Seyfert galaxies. The 
major absorption component is occuring against the bright radio 
knot offset a few hundred parsecs from the core. While there are 
indication that the absorbing material is associated with the large 
scale \HI\ disk, it is clearly (and violently) disturbed by the
passage of the jet. We suspect that more sensitive observations 
may reveal similar absorption profiles in other Seyfert galaxies 
with fainter radio sources.

\acknowledgements
We wish to thank the referee, Jack Gallimore, for his useful comments.

\newpage

\figcaption{ATCA 3 cm radio continuum image from Morganti et al.\ (1998).}
\figcaption{VLBI 13 cm radio continuum image of (the western lobe of)  IC~5063}
\figcaption{VLBI \HI\ absorption profile converted in optical depth.
For comparison, the ATCA profile is superimposed as dashed line.}

\figcaption{VLBI (solid line)  and ATCA (dashed line) \HI\ absorption profile 
normalised to the same optical depth of the main absorption component.}

\figcaption{Schematic diagram of the model presented in \S 5 to explain the 
characteristics of the \HI\ absorption observed in the western lobe of
IC~5063. See text for details.}

%
\begin{figure}
\centerline{\psfig{figure=oosterloo.fig1.ps,width=10cm,angle=-90}}
{\bf Fig.1}
\end{figure}

\newpage
%
\begin{figure}
\centerline{\psfig{figure=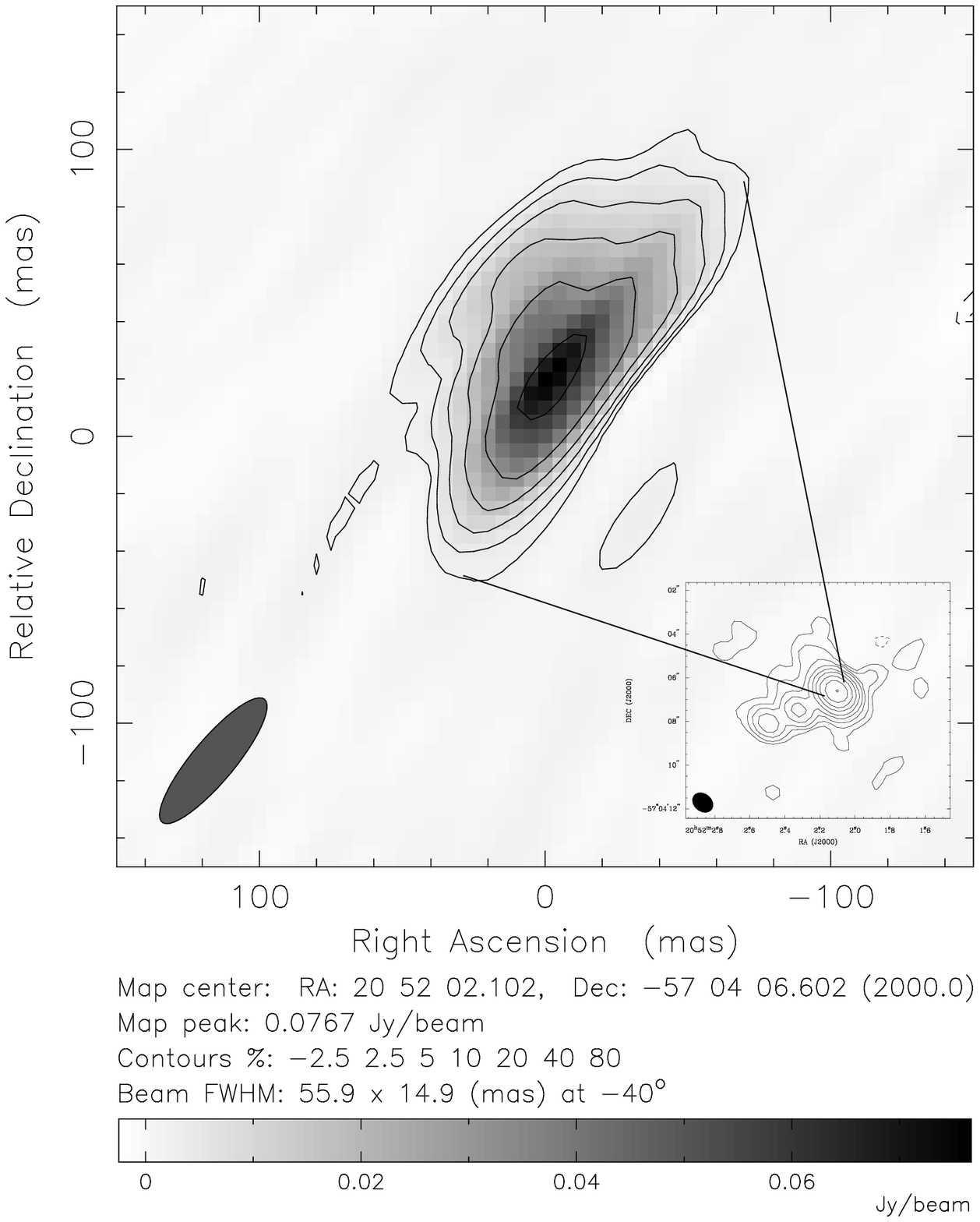,width=9cm,angle=0}}
{\bf Fig.2}
\end{figure}

\begin{figure}
\centerline{\psfig{figure=oosterloo.fig3.eps,width=8.5cm,angle=-90}}
{\bf Fig.3}
\end{figure}

\begin{figure}
\centerline{\psfig{figure=oosterloo.fig4.eps,width=8.5cm,angle=-90}}
{\bf Fig.4}
\end{figure}

\begin{figure}
\centerline{\psfig{figure=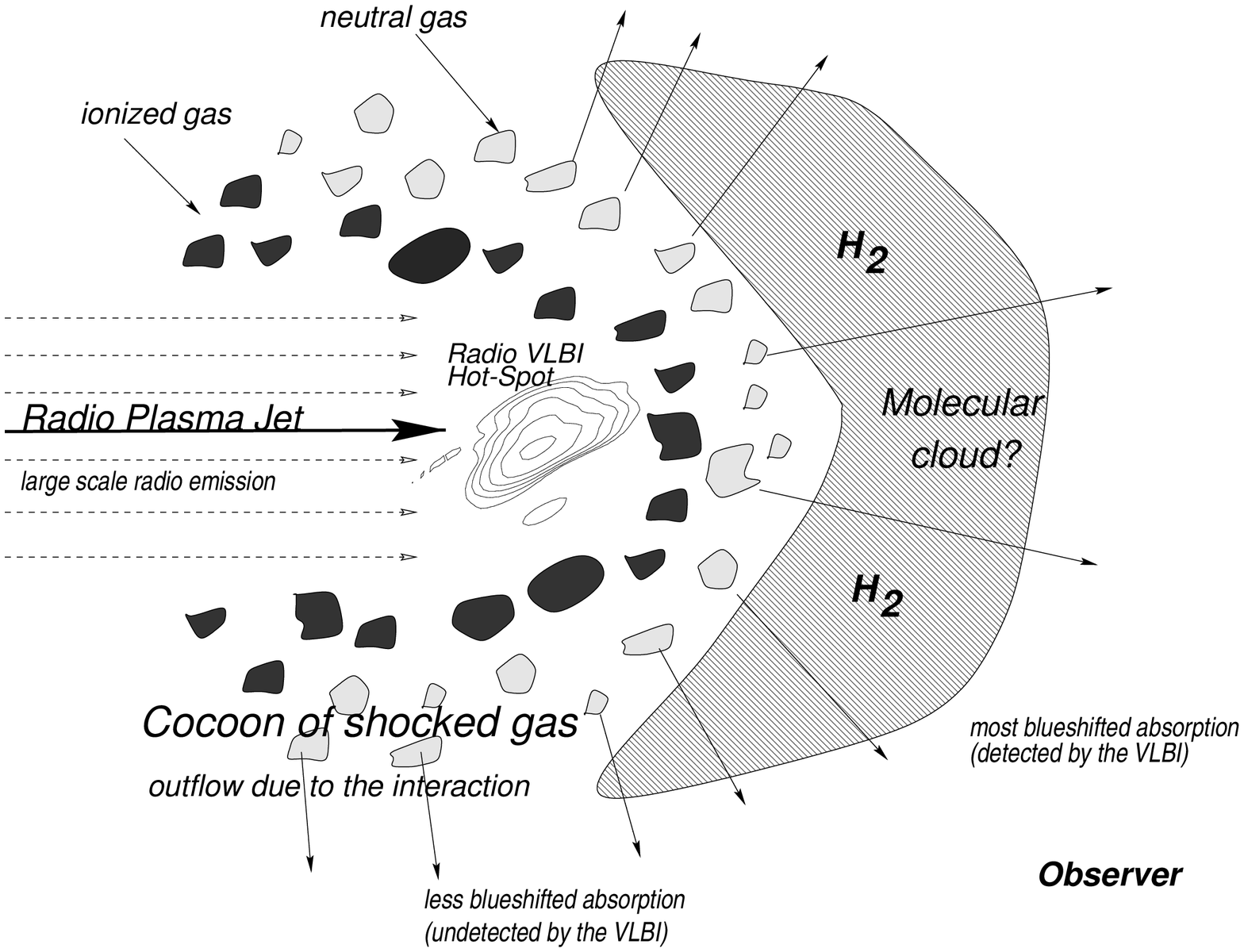,width=13cm}}
{\bf Fig.5}
\end{figure}


\newpage

\begin{thebibliography}{}
         
\bibitem{} Aoki K., Ohtani H., Yoshida M. \& Kosugi G. 1996, AJ 111, 140
\bibitem{} Bahcall J.N., Ekers R.D. 1969, ApJ 157, 1055
\bibitem{} Bicknell G.V., Dopita M. A., Tsvetanov Z.I., Sutherland R. S.
1998, ApJ 495, 680
\bibitem{} Bransford M.A., Appleton P.N., Heisler C.A., Norris R.P., Marston
A.P. 1998, ApJ 497, 133
\bibitem{} Brinks, E., \& Mundell, C.~G.  1996, in {\it The Minnesota
Lectures on Extragalactic Neutral Hydrogen}, ed.\ E.~D.\ Skillman, ASP
Conf.\ Series, 106, 268 
\bibitem{} Capetti A., Macchetto F., Axon D.J., Sparks W.B. \& Boksenberg
A. 1996, ApJ 469, 554
\bibitem{} Cole G.H.J., Pedlar A., Mundell C.G.,
Gallimore J.F.  \& Holloway 1998, MNRAS 301,782 
\bibitem{} Colina L., Sparks W.B. \& Macchetto F. 1991, ApJ 370, 102
\bibitem{} Draine B.T., Roberge W.G. \& Dalgano A. 1983, ApJ 264, 485
\bibitem{} Falcke H., Wilson A.S., Simpson C.  1998, ApJ 502, 199
\bibitem{} Gallimore J.F., Holloway A.J., Pedlar A.,
Mundell C.G.  1998, A\&A 333, 13 
\bibitem{} Gallimore J.F., Baum S.A., O`Dea C.P., Pedlar A., Brinks E. 1999,
ApJ in press (astro-ph/9905267)
\bibitem{} Koyama, K., Awaki, H., Iwasawa, K., \& Ward, M. J. 1992, ApJ, 399, L129
\bibitem{} Kulkarni V.P.  et al.  1998, ApJ 492, L121 
\bibitem{} Morganti R., Oosterloo T.  \& Tsvetanov Z., 1998, AJ, 115, 915 (M98)
\bibitem{} Pedlar A., Howley P., Axon D.J.  \& Unger
S.W., 1992, MNRAS, 259, 369 
\bibitem{} Mundel C.G., Pedlar A., Baum
S.A., O'Dea C.P., Gallimore J.F.  \& Brinks E.  1995, MNRAS 272, 355
\bibitem{} Roy A.L., Colbert E.J.M., Wilson A.S., Ulvestad J.S. 1998, ApJ 504,
147
\bibitem{} Roy A.L. \& Norris R.P. 1997, MNRAS 289, 824
\bibitem{} Sault R.J., Teuben P.J., Wright M.C.H. 1995, in {\it ``Astronomical 
Data Analysis Software and Systems IV''}, eds. R. Shaw, H.E. Payne and
J.J.E. haynes, ASP Conf. Series, 77, 433 
\bibitem{} Shepherd M.C. 1997, in ``Astronomical Data Analysis Software and
Systems IV'', ASP Conf. Series Vol. 125, Hunt G. \& Payne H.E. (eds.), p.77
\bibitem{} Sternberg A. \& Dalgano A. 1989, ApJ 338, 197
\bibitem{} Wagner S.J. \& Appenzeller I., 1989, A\&A, 225, L13
\bibitem{} Wiklind T., Combes F. \& Henkel C. 1995, A\&A 297, 643
\bibitem{} Wilson A.S. 1997, in {\it ``Emission lines in Active Galaxies: 
new methods and techniques''}, eds.\ B.M.  Peterson, F.-Z.\ Cheng and A.S.\
Wilson, ASP Conf.\ Series Vol.\ 113, p.\ 264

\end{thebibliography}
\end{document}